%% file: OpenFMM_arxiv.tex
\documentclass[10pt,a4paper,twocolumn]{revtex4-1}
\usepackage{graphicx} 
\usepackage{amsmath} 
\usepackage{amsfonts}
\usepackage{bm}	

\begin{document}

\title{Open geometry Fourier modal method: Modeling nanophotonic structures in infinite domains}

\author{Teppo~H\"ayrynen} \email{tepha@fotonik.dtu.dk}
\author{Jakob~Rosenkrantz~de~Lasson}
\author{Niels~Gregersen}

\affiliation{DTU Fotonik, Department of Photonics Engineering, Technical University of Denmark, \O rsteds Plads, Building 343, DK-2800 Kongens Lyngby, Denmark}

\date{\today}

%\ociscodes{
%(050.1755) Computational electromagnetic methods; 
%(350.3950)   Micro-optics;
%(230.7370)   Waveguides
%(000.3860) Mathematical methods in physics; 
%(000.4430) Numerical approximation and analysis; 
%}
%\doi{\url{http://dx.doi.org/10.1364/josaa.XX.XXXXXX}}

% Packages
%\usepackage[dvips]{graphicx} % ps-figures
%\usepackage[english]{babel}

%\usepackage{graphicx} 
%\usepackage{epstopdf} % converts eps->pdf so that article can include both the eps and the pdf figures

% New commands
\newcommand{\ci}{\mathrm{i}}
\newcommand{\vekn}[1]{\boldsymbol{\mathrm{#1}}}  	% Bold face and non-italized
\newcommand{\vekE}{\vekn{E}} % Vector E
\newcommand{\vekH}{\vekn{H}} % Vector E
\newcommand{\ddd}{\:\mathrm{d}}						% Differential d with space before
\newcommand{\iim}{\mathrm{i}}	% imaginary symbol
\newcommand{\ud}{\mathrm{d}}	% differantial d 
\newcommand{\kommentti}[1]{}

\begin{abstract}
We present an open geometry Fourier modal method based on a new combination of open boundary conditions and an efficient $k$-space discretization. The open boundary of the computational domain is obtained using basis functions that expand the whole space, and the integrals subsequently appearing due to the continuous nature of the radiation modes are handled using a discretization based on non-uniform sampling of the $k$-space. We apply the method to a variety of photonic structures and demonstrate that our method leads to significantly improved convergence with respect to the number of degrees of freedom, which may pave the way for more accurate and efficient modeling of open nanophotonic structures.
\end{abstract}

%\begin{document}

\maketitle

\section{Introduction}

Many important properties of photonic structures, like cavities~\cite{Vahala2003} and waveguides~\cite{Lecamp2007,MangaRao2007}, depend on their radiative losses that stem from coupling of energy into freely propagating optical modes that escape the system. The quality, or $Q$, factor of photonic resonators as well as the spontaneous emission (SE) $\beta$ factor in waveguides are important figures of merit, for example in the analysis of nanolasers~\cite{Strauf2011} and single-photon sources~\cite{Gregersen2013}, and these quantities depend sensitively on the radiative losses. In modeling such open photonic systems, the choice of boundary conditions (BCs) at the computational domain edges becomes crucial and may impact the results not just quantitatively, but also qualitatively. Integral equation-Green's function formulations inherently adopt this openness~\cite{Lavrinenko2014Chap7},
while for numerical techniques relying on finite-sized computational domains, like the finite-difference time-domain (FDTD) method~\cite{Taflove1995} and the finite element method (FEM)~\cite{Reddy2005}, this is achieved using artificial absorbing boundaries, typically in the form of so-called perfectly matched layers (PMLs)~\cite{Berenger1994}.

In Fourier-based modal expansion techniques~\cite{Noponen1994,Moharam1995}, PMLs can be implemented using complex coordinate transforms~\cite{Hugonin2005}. The absorbing boundaries are implemented by mapping the real spatial coordinates into complex ones, which is straightforward to implement. In turn, it is unclear which complex coordinate transform to implement and why, and there have been no systematic studies on the influence of PML parameters and the size of the computational domain on computed quantities of interest. \textit{In addition} to Fourier resolution convergence checks, the size of the computational domain should be varied to estimate the computational accuracy, but this is rarely done~\cite{Gregersen2010,Pisarenco2010,Kristensen:12}. 
In our experience, different choices of PML parameters and domain sizes lead to results that agree qualitatively, but that may vary substantially,
for example, the errors of $Q$ factors $\sim 20\%$ \cite{Gregersen2010} and the errors of dipole coupling to radiation modes  $\sim 15-25\%$ \cite{deLassonThesis} have been reported.

Instead of searching an extremely large PML parameter space without intuitive or clear guidelines, we here propose a different technique that relies on finite-sized structures and open BCs, with fields expanded via Fourier \textit{integrals} instead of Fourier \textit{series}. The use of Fourier integrals, in principle, gives an exact description, but for numerical implementation a $k$-space discretization is required, that we, however, have the freedom to choose. Similar ideas have previously been reported for two-dimensional (2D)~\cite{Guizal2003} and rotationally symmetric three-dimensional (3D)~\cite{Bonod2005, Bava01, Dems10} structures, but without discussion of the important problem of choosing the $k$-space discretization. Furthermore, the important example of dipole emission, that depends sensitively on the proper implementation of the open BCs, was not treated in these works. In this manuscript, we address both these central issues. 
Our examples include calculations of light emission from emitters placed in rotationally symmetric waveguides \cite{Claudon2013} and reflection of the fundamental mode from a waveguide-metal interface \cite{Friedler08}.
We term this new approach open geometry Fourier modal method (oFMM).

This manuscript is organized as follows. Section \ref{sec:theory} outlines the theory of the oFMM approach while the details are given in appendix \ref{app:rotsym}. The details of the new discretization scheme is discussed in Sec. \ref{sec:disc}. The method is tested in several structures by calculating dipole emission rates, $\beta$ factors and modal reflection coefficients in Sec. \ref{sec:results}.

\section{Theory} \label{sec:theory}

In this section we outline the derivation of the open BC formalism and introduce the theoretical concepts required to understand the results of the following sections. The detailed derivations of the open BCs formalisms in rotationally symmetric geometry is given in appendix \ref{app:rotsym}.

\subsection{Open boundary condition formalism}

We employ the open BCs formalism to describe the electromagnetic field propagation in rotationally symmetric structures. Complete vectorial description is used in connection with Fourier expansion to describe the Maxwell's equations in a $z$-invariant material section. Using cylindrical coordinates in the rotationally symmetric case allows for simplification of the problem to 1D expansion in the radial coordinate. The $z$-dependence is treated by combining $z$-invariant sections using the scattering matrix formalism (see, e.g., \cite{Li96_josa13_5} and \cite{Lavrinenko2014Chap6} for details).
Our task is then to determine the lateral electric and magnetic field components of the eigenmodes, which are subsequently used as an expansion basis for the optical field. In the conventional FMM, this is done by expanding the field components as well as the permittivity $\varepsilon(\mathbf{r}_{\perp})$ and $\eta(\mathbf{r}_{\perp})\equiv 1/\varepsilon(\mathbf{r}_{\perp})$ in Fourier \textit{series} in the lateral coordinates $\mathbf{r}_{\perp}$ on a \textit{finite-sized} computational domain, implying that these functions vary \textit{periodically} in these coordinates.  In the open boundary formalism, we instead consider an \textit{infinite-sized} computational domain and employ expansions in Fourier \textit{integrals}. Our approach describing rotationally symmetric structures uses the Bessel $J$-functions as basis functions, as first proposed in \cite{Bonod2005}. In the following, we describe the general steps and equations required to expand the field components and to solve for the expansion and propagation coefficients. The specific equations and derivations are given in the appendix and referenced throughout this section.

Starting from the time-harmonic Maxwell's equations $\nabla \times \mathbf{E}(\mathbf{r}) = \iim\omega\mu_0 \mathbf{H}(\mathbf{r}) $
and $\nabla \times \mathbf{H}(\mathbf{r}) = -\iim\omega \varepsilon(\mathbf{r}) \mathbf{E}(\mathbf{r}) $ 
(written using cylindrical coordinates in Eqs. (\ref{eq:me_Et})--(\ref{eq:me_Ez}) in the appendix), where $\varepsilon$ is the permittivity of the medium, $\mu_0$ is the vacuum permeability, $\omega$ is the angular frequency and $\mathbf{E}$ and $\mathbf{H}$ the electric and magnetic fields, 
we obtain 
\begin{equation}
\nabla \times [\nabla \times \mathbf{E}(\mathbf{r})] = \omega^2\mu_0 \varepsilon(\mathbf{r}) \mathbf{E}(\mathbf{r}),
%\label{eq:}
\end{equation}
which is given in cylindrical coordinates in Eqs. (\ref{eq:BHrc})--(\ref{eq:BHzc}).
The fields in single $z$-invariant section can be expanded using the eigenmodes of the system as
%\begin{equation}
\begin{align}
\nonumber
\mathbf{E}(\mathbf{r}_{\perp}, z) &= \sum_{j} a_j \mathbf{E}_j(\mathbf{r}_{\perp})\exp(\pm\iim \beta_j z) \\
&+ \int a(k) \mathbf{E}(k,\mathbf{r}_{\perp})\exp(\pm\iim \beta(k) z)\ud k, 
\label{eq:zprop}
\end{align}
%\end{equation}
where $\beta_j$ and $\beta(k)$ denote the propagation constants, and $a_j$ and $a(k)$ the weights of the corresponding modes. Furthermore, the summation index $j$ denotes all the guided modes while the integral accounts for the radiation and evanescent modes. In numerical simulations the continuous integral is approximated by a sum as
\begin{eqnarray}
\nonumber
&& \int a(k) \mathbf{E}(k,\mathbf{r}_{\perp})\exp(\pm\iim \beta(k) z)\ud k \\
&\approx& \sum_l a_l \mathbf{E}_j(k_l,\mathbf{r}_{\perp})\exp(\pm\iim \beta_l z) \Delta k_l,
\label{eq:rad_mod_disc}
\end{eqnarray}   
where $\Delta k_l = k_l - k_{l-1}$ and $k_{l} = \sqrt{(nk_0)^2-\beta_{l}^2}$, with $k_0$ denoting the wavenumber in vacuum and $n$ being the refractive index of the material. Similar equations hold for the magnetic field.

Using the discretized eigenfunction expansion in Eqs.  (\ref{eq:zprop})--(\ref{eq:rad_mod_disc}), the fields in each $z$-invariant section can be expressed with column vectors $\mathbf{a}$ consisting of electric and magnetic field expansion coefficients, $[a_j, a_l\Delta k_l]^{\mathrm{T}}$, all denoted with the single index $j$ in the following.
Thus, taking the $z$ derivative of Eq. (\ref{eq:zprop}) we can formulate an eigenvalue problem describing the fields in the system as 
\begin{eqnarray}
\mathbf{M}\mathbf{a} = \iim\beta\mathbf{a}, 
\end{eqnarray}    
where the elements of matrix $\mathbf{M}$ are obtained by expanding the eigenfunction in Fourier-Bessel basis in rotationally symmetric geometry as discussed below.

Since the eigenfunctions are specific to each layer, we choose a general basis and expand the eigenfunctions in each layer using the common basis. Thus any function (the field components and the relative permittivity) can be expanded as a Fourier transform
\begin{equation}
f(\mathbf{r}_{\perp}) = \int_{k_{\perp}}c_f(k_{\perp})g(k_{\perp}, \mathbf{r}_{\perp}) \ud k_{\perp},
\label{eq:F_exp}
\end{equation}
where $k_{\perp}$ is transverse wavenumber  
while $c_f(k_{\perp})$ and $g(k_{\perp}, \mathbf{r}_{\perp})$ are the expansion coefficients  and the basis functions.
In the rotationally symmetric case $k_{\perp}=k_r$ and $\mathbf{r}_{\perp}=r$, and the basis functions $g(k_{r}, r)$ are the Bessel $J$-functions (cf. Eqs. (\ref{eq:Er_exp})--(\ref{eq:Et_exp})). 
While in the analytical definition of the Fourier transform the expansion basis in integral (\ref{eq:F_exp}) is infinite,  for the numerical calculations the basis must be truncated as 
\begin{eqnarray}
\nonumber
&&  \int_{k_{\perp}}  c_f(k_{\perp})g(k_{\perp}, \mathbf{r}_{\perp}) \ud k_{\perp} \\
&\simeq&
\sum_{m=1}^{M} c_f(k_{\perp,m})g(k_{\perp,m}, \mathbf{r}_{\perp}) \Delta k_{\perp,m},
\label{eq:f_exp_trunc}
\end{eqnarray}   
where the discretization steps $\Delta k_{\perp,m}$ will be a function of the index $m$ in the generalized approach as will be discussed in section \ref{sec:disc}. This is in contrast to previous approaches \cite{Guizal2003, Bonod2005}, and we show later that such a non-uniform discretization is a significant improvement.
The expansions in cylindrical coordinate system are given by Eqs. (\ref{eq:Er_exp})--(\ref{eq:Et_exp}).
Furthermore, the elements of $\mathbf{M}$ are given in Eqs. (\ref{eq:bE})--(\ref{eq:cH}). Solving the eigenvectors and eigenvalues of matrix $\mathbf{M}$ yields the expansion coefficients and propagation constants in the $z$-invariant structures, while
the fields in the full structure are then obtained by combining the $z$-invariant sections using the scattering matrix formalism.

\subsection{Dipole emission} \label{sec:emission}

The field emitted by a point dipole placed at $\mathbf{r}_{\mathrm{pd}}$ inside a $z$-invariant structure can be represented as
\begin{eqnarray}
%\mathbf{E}(\mathbf{r}) = \sum_{m=1}^{M} a_m \mathbf{E}_m(\mathbf{r}),
\nonumber
\mathbf{E}(\mathbf{r}) &=& \sum_{j}a_j(\mathbf{r}_{\mathrm{pd}},\mathbf{p}) \mathbf{E}_j(\mathbf{r}) \\ 
                       &=& \sum_{j}\sum_{m} a_j c_{j,m} \mathbf{g}_{m}(\mathbf{r})\Delta k_{\perp,m}e^{\pm\iim\beta_j (z-z_{\mathrm{pd}})},
\label{eq:em_Ef}
\end{eqnarray}
where $\mathbf{E}_j(\mathbf{r})$ is the electric field of $j$th eigenmode, and $a_j(\mathbf{r}_{\mathrm{pd}},\mathbf{p})$ is the dipole coupling coefficient to mode $j$, which can be calculated using the Lorentz reciprocity theorem \cite{Lavrinenko2014Chap6}. The coupling coefficient depends on the dipole position $\mathbf{r}_{\mathrm{pd}}$ and dipole moment $\mathbf{p}$ through a dot-product $\mathbf{p} \cdot \mathbf{E}_j(\mathbf{r}_{\mathrm{pd}})$. For the sake of notational clarity we we omit these dependencies in the following.  Furthermore, $c_{j,m}$ are the expansion coefficients for mode $j$, and $\mathbf{g}_{m}(\mathbf{r}_{\perp})$ are the basis functions. 

The emitted field (\ref{eq:em_Ef}) consists of three contributions \cite{Snyder}: guided modes, radiating modes, and evanescent modes. In a waveguide surrounded by air, the modes are guided if the propagation constant $\beta_j$ obeys $k_0^2 < \beta_j^2 \le (n_w k_0)^2$, where $n_w$ is the refractive index of the waveguide. In contrast the modes are radiating if $0<\beta_j^2 \le k_0^2$, and evanescent if $\beta_j^2 < 0$. We will apply this classification in section \ref{sec:disc} when we investigate discretization schemes.

The normalized power emitted by dipole to a selected mode can be expressed as  \cite{Novotny2012Chap8, Lavrinenko2014Chap6}
\begin{eqnarray}
\nonumber
\frac{P_j}{P_{\mathrm{Bulk}}} &=& \frac{\mathrm{Im}\{ a_j \mathbf{E}_j(\mathbf{r}_{\mathrm{pd}})  \}}{P_{\mathrm{Bulk}}} \\
 &=& \frac{\mathrm{Im}\{ \sum_m a_j c_{j,m} \mathbf{g}_{m}(\mathbf{r}_{\mathrm{pd}})\Delta k_{\perp,m}  \}}{P_{\mathrm{Bulk}}}, 
%\label{eq:}
\end{eqnarray}
where $P_{\mathrm{Bulk}}$ is the emitted power in a bulk medium. The normalized emitted power is equal to the normalized emission rate \cite{Novotny2012Chap8} $\gamma_j/\gamma_{\mathrm{Bulk}} = P_j/P_{\mathrm{Bulk}}$, where $\gamma_j$ and $\gamma_{\mathrm{Bulk}}$ are the emission rates to mode $j$ and in a bulk, respectively. In the following we will use only the normalized unitless quantity $\Gamma_j = \gamma_j/\gamma_{\mathrm{Bulk}}$ for the emission rate.
Thus, the spontaneous emission factor (i.e. the $\beta$ factor), defined as the ratio of emission to the fundamental mode (FM) and the total emission \cite{Claudon2013}, is obtained as 
\begin{equation}
%\beta = \frac{\Gamma_{\mathrm{HE}_{11}}}{ \Gamma_{\mathrm{tot}} } 
%
\beta 
= \frac{a_{\mathrm{FM}} \mathbf{E}_{\mathrm{FM}}(\mathbf{r}_{\mathrm{pd}})}{ \sum_{j} a_j \mathbf{E}_j(\mathbf{r}_{\mathrm{pd}}) }
= \frac{a_{\mathrm{FM}} \sum_m  c_{\mathrm{FM},m} \mathbf{g}_{m}(\mathbf{r}_{\mathrm{pd}})\Delta k_{\perp,m} }{ \sum_{j}\sum_{m} a_j c_{j,m} \mathbf{g}_{m}(\mathbf{r}_{\mathrm{pd}})\Delta k_{\perp,m} }. 
%\label{eq:}
\end{equation}

\section{Discretization scheme} \label{sec:disc}

In addition to the open BCs described in the previous section, an advantage of the presented method is that it enables using a non-uniform $k$-space discretization, which allows having a high cut-off value together with dense sampling $k$-space regions and still maintaining moderate total number of modes, i.e. achieving the required accuracy with less computational power.
In this section, we investigate how to select the cutoff value $k_{\mathrm{cut-off}}$ and how to sample the $k$-space effectively. The numerical tests in section \ref{sec:results} show that faster convergence is achieved using an appropriate mode sampling scheme.  

The transverse wavenumber values in the conventional modal expansion approach \cite{Noponen1994}  are selected equidistantly 
\begin{equation}
k_m = m \Delta k = \frac{m}{M+1} k_{\mathrm{cut-off}},
%\label{eq:}
\end{equation} 
where $m=1\dots M$ and the discretization step size depends on the selected cut-off value $k_{\mathrm{cut-off}}$ and number of modes $M$ as $\Delta k = k_{\mathrm{cut-off}}/(M+1)$. 

\begin{figure}[h]
\begin{center}
\includegraphics[width=6cm]
%{figures/k_discretization.eps}%
{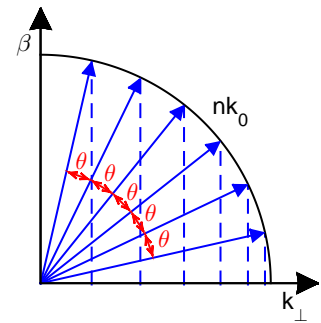}%
\caption{Non-uniform discretization scheme:
In a bulk medium all propagation directions have equal weights. Therefore, the wavevector $\mathbf{k}$ is sampled in the $(\beta, \mathbf{k}_{\perp})$-plane using equidistant angles as shown by $\theta$ in the figure.
Due to the uniform angle distribution, the $k_{\perp}$ discretization is more dense close to $nk_0$.
% Figure inspired by \cite{deLasson2012}.       
\label{fig:k_discr}}
\end{center}
\end{figure}

In bulk, light emission occurs with equal weights in all directions. Therefore, a natural starting point for the discretization scheme is to sample the wavevectors in the $(\beta, \mathbf{k}_{\perp})$ plane with equidistant angles \cite{deLasson2012}, as shown in Fig. \ref{fig:k_discr}. Then the different transverse wavenumber values are given by $k_m = nk_0\sin(\theta_m)$, where the equidistantly sampled angles $0<\theta_m < \pi/2$ are measured from the $\beta$ axis. Although the values of $\theta_m$ are selected uniformly, the values of $k_m$ are more densely sampled in the proximity of $nk_0$, cf. Fig. \ref{fig:k_discr}. If, instead of a bulk, we consider a structure like a nanowire consisting of several materials it is necessary also to account for the modes beyond $nk_0$.

\begin{figure}[h]
\begin{center}
\includegraphics[width=\columnwidth]%
%{figures/Emission_components_1500_10.eps}%
{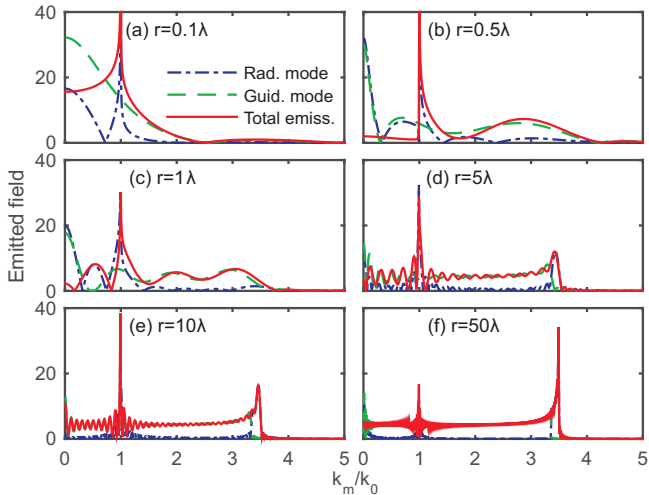}%
\caption{The Fourier components of a point dipole emission defined in Eq. (\ref{eq:Em_comp}).
The figures show the calculated radiation and guided mode contributions and the total emission as function of the radial wave number in $z$-invariant nanowires of varying radius. The nanowire has a refractive index of $n_w=3.5$ and the wavelength is $\lambda=950$ nm. An equidistant $k$-discretization with 1500 points and $k_{\mathrm{max}} = 10k_0$ was used.
\label{fig:Em_comp}}
\end{center}
\end{figure}

To obtain insight into the discretization in different types of structures, we first investigate emission from a radially oriented point dipole placed on the axis of rotationally symmetric infinite semiconductor nanowires having radius from sub-wavelength to several wavelengths and a refractive index $n_w$, see Fig \ref{fig:Em_comp}. The radial component of the emitted electric field $E_r(r) = \sum_{j} a_j E_{r,j}(r) = \sum_{j} a_j \sum_{m} c_m g_{r,m}(r)\Delta k_m$ can be written as follows (cf. Sec. \ref{sec:theory} \ref{sec:emission}, Eqs. (\ref{eq:f_exp_trunc}) and (\ref{eq:em_Ef}) and Appendix \ref{app:rotsym}) by rearranging the terms 
\begin{eqnarray}
\nonumber
&& E_r(r) = \iim \sum_m \Bigg[ \\
&&  \sum_{j=\mathrm{g.~m.}} a_j \mathcal{E}_{j,m}(r) + \sum_{j=\mathrm{r.~m.}} a_j 
\mathcal{E}_{j,m}(r) \Bigg]k_m\Delta k_m 
\label{eq:Em_comp}
\end{eqnarray}
where a short hand notation $\mathcal{E}_{j,m}(r) = b^E_{n,m,j}J_{n+1}(k_mr) - c^E_{n,m,j}J_{n-1}(k_mr)$ has been used for the radial component of the electric field defined in Eq. (\ref{eq:Er_exp}). The sum over $j$ describes the modes of the structure, while index $m$ accounts for the expansion of the modes using the selected basis functions. The first summation inside the brackets in Eq. (\ref{eq:Em_comp}) describes the guided mode ($k_0^2<\beta_j^2 \le (n_wk_0)^2$) contribution, while the second summation describes the radiation mode ($0<\beta_j^2 \le k_0^2$) contribution to the total emission with radial wavenumber $k_m$. Figure \ref{fig:Em_comp} shows the guided and radiation mode contribution and the their sum as functions of radial wavenumber.
The emitted electric field has a peak around $k_m = k_0$. When the radius increases, also a peak around $k_m = n_wk_0$ gradually builds up, while in the bulk limit ($r/\lambda\gg 1$) the peak around $k_0$ disappears. These results indicate that (i) for wires with radius $r \lesssim \lambda$ the $k$-space should be densely sampled around $k_0$, (ii) for wider structures dense sampling around $n_wk_0$ is also required, while (iii) in bulk dense sampling is required only around $n_wk_0$. 
Thus, since we are mainly interested in the region where the wire radius is of the order of or smaller than the wavelength, we will use the following discretization scheme that is dense and symmetric around $k_0$ and where the discretization step-size gradually increases towards the cut-off value.
Let $M_1, M_2, M_3$ be the fixed number of $k$ values on the intervals $(0,k_0)$, $(k_0, 2k_0)$, and $(2k_0, k_{\mathrm{cut-off}})$, respectively. Then we can write
\begin{widetext}
\begin{equation}
\begin{array}{ll}
k_m^{(1)} = k_0\sin(\theta_m), ~~\theta_m = \frac{\pi}{2}\frac{m}{M_1+1}, & m=1,\dots, M_1 \\ 
k_m^{(2)} = k_0[2-\sin(\theta_m)],~~  \theta_m = \frac{\pi}{2}\left(1+\frac{m}{M_2+1}\right), & m=1,\dots, M_2 \\ 
k_m^{(3)} =  k_{n_2}^{(2)} + \delta_1 m  + \frac{\delta_2}{2} m(m+1),  &  m=1,\dots, M_3, 
\end{array}
\end{equation}
\end{widetext}
where we use a symmetric dense sampling around $k_0$ by setting $M_2=M_1$. Furthermore,
$\delta_1 = \Delta k_{M_2}^{(2)}$ is then biggest step size in the symmetric region, and 
$\delta_2 = 2[k_{\mathrm{cut-off}}-k_{M_2}^{(2)}-M_3\delta_1]/[M_3(M_3+1)]$. When modeling bulk materials, $k_0$ will be multiplied with the refractive index as concluded above.    
In the following we will use $M_1=M_2=M_3 = M/3$. The optimal values of $M_i$ may vary depending on the geometry, but this choice limits the number of free parameters to the total number of modes $M$ and to the cut-off value $k_{\mathrm{cut-off}}$. Our examples will show that this selection leads to faster convergence of the calculations than using the equidistant discretization scheme.

In the next section, we use these discretization schemes in modeling of various structures and compare the convergence and required computational power with those obtained using conventional discretization scheme. When comparing the different discretization schemes, we use the same cut-off value and the same number of modes for both of the schemes.

\section{Results and discussions} \label{sec:results}

Next, after introducing the principles of open BC formalism together with the new discretization strategy, we are ready to test the method with several numerical examples. 
The purpose of these selected examples is to show that the calculations using oFMM formalism converge towards a well-defined open geometry limit and that faster convergence can be achieved using the discretization schemes introduced in Sec. \ref{sec:disc} compared to using the conventional equidistant discretization. 
We start with calculating the dipole emission rates (or emission power) in bulk and close to an interface since these results can be verified analytically. After these basic checks, we investigate the performance of our method for the cases of light emission from emitters in waveguides as well as the case of reflection at a waveguide-metal interface, all of which depend critically on a correct and accurate description of the open boundaries.

\subsection{Dipole emission in bulk and close to an interface}

\begin{figure}[h]
\begin{center}
\includegraphics%[width=8cm]
[height=9.0cm]
%{figures/bulk_and_interface_fig.eps}%
{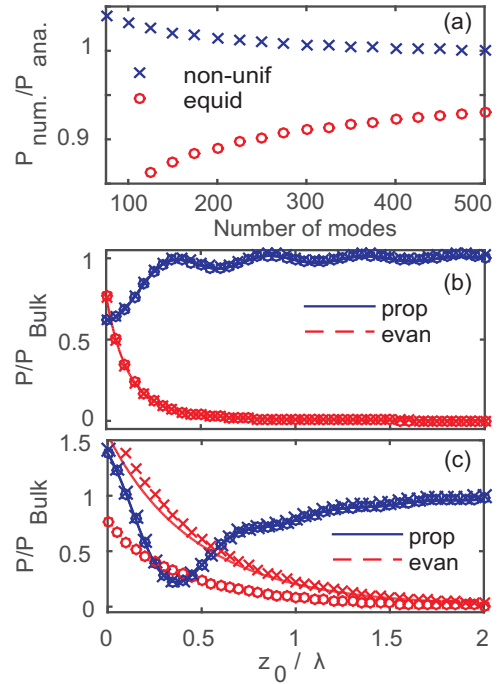}%
\caption{
(a)
 Calculated emission power $P_{\mathrm{num}}$ in a bulk ($n_b=1$) normalized with analytical result $P_{\mathrm{ana}}$ with a fixed wavelength $\lambda = 950$ nm. 
\textit{Crosses:} Numerical result calculated using the non-uniform discretization scheme. 
\textit{Circles:} Numerical result calculated using the equidistant discretization scheme. 
Both numerical schemes have the wavenumber cut-off value $n_bk_0$ in the bulk and the horizontal axis shows the number of modes included in the calculations.  
(b) Normalized dipole emission power in air in front of glass ($\varepsilon = 2.25$) half-space. The dipole is parallel to the interface.
(c)
Normalized power emitted by point dipole placed in air close to an air-metal interface ($\varepsilon=-41+2.5\iim$).
The dipole is perpendicular to the interface.
\textit{Lines:} Semi-analytical result. 
\textit{Crosses:} Numerical result using 200 modes with non-uniform discretization.
\textit{Circles:} Numerical result using 200 modes with equidistant discretization.
Numerical results are calculated using  a cut-off value of $2k_0$. The powers are normalized with the bulk value and the distance $z_0$ from the interface with the wavelength $\lambda = 950$nm. 
\label{fig:bulk_emission}}
\end{center}
\end{figure}

As a first example, we consider dipole emission in bulk and close to a bulk-bulk interface. Both of these examples can also be solved analytically \cite{Novotny2012Chap8, Novotny2012Chap10} allowing easy comparison of the convergence of the results. Figure \ref{fig:bulk_emission}(a) shows the dipole emission power in a bulk material ($n_b=1$) calculated using the rotationally symmetric model and normalized with the analytical result. Numerical results are calculated using both the equidistant discretization and the non-uniform discretization presented in Sec. \ref{sec:disc}. The obtained results show that applying the non-uniform discretization leads to much faster convergence of the emission rates.

In the bulk case, only propagating modes contribute to the light emission and the emission rate converges provided that enough propagating modes are included in the calculation. In contrast, in the case of a dipole emitter in close proximity of an interface also the evanescent modes contribute through evanescent mode scattering at the interface and re-excitation of the propagating modes. As a next example we, therefore, investigate the interface case.  

Figure \ref{fig:bulk_emission}(b) shows the power emitted by a dipole close to an air-glass interface
while Fig. \ref{fig:bulk_emission}(c)  shows  that for a dipole close to an air-metal interface.
The values of the metal and glass permittivities are $\varepsilon=-41+2.5\iim$ and $\varepsilon=2.25$, respectively. In contrast to the bulk case in Fig. \ref{fig:bulk_emission}(a)  where the cut-off was $n_bk_0$ ($n_b=1$), we now need to include the evanescent modes. 
Figures \ref{fig:bulk_emission}(b) and (c) show the separate contributions from propagating and evanescent modes to the emission rate. Again, the non-uniform discretization leads to faster convergence, especially for the contribution from the evanescent modes.

\subsection{Dipole emitter in a rotationally symmetric waveguide}

Next we investigate the emission in waveguides by calculating the emission rates to selected modes and the spontaneous emission factor $\beta$.  In contrast to the bulk and interface cases investigated in the previous section, we face an additional computational challenge which is to compute the radiating modes accurately. The waveguides considered in nanophotonics usually support only a few guided modes. However, the total emission rate and thus the $\beta$ factor depend on emission to continuum of radiation modes that can radiate light out of the waveguide. Calculating the radiation modes accurately requires more extensive calculations than the emission on bulk as will be seen in the following examples.

\begin{figure}[h]
\begin{center}
\includegraphics{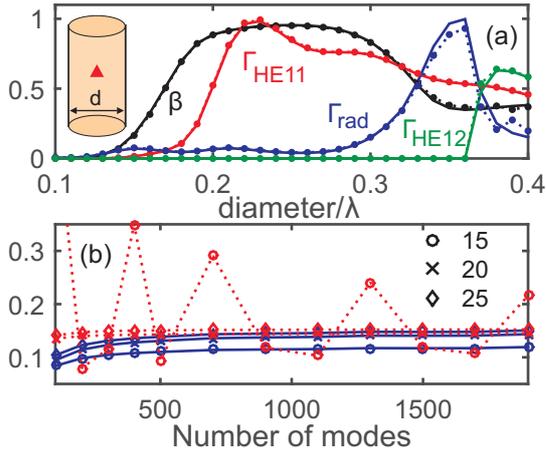}
\caption{
Emission from a point dipole placed on the axis of a infinitely long rotationally symmetric nanowire of diameter $d$. (a) The $\beta$ factor and normalized emission rates to the first and second guided modes $\mathrm{HE}_{11}$, $\mathrm{HE}_{12}$ and radiation modes as functions of $d$. The nanowire refractive index is $n=3.45$ and the wavelength is $\lambda=950$ nm. 
The lines show results obtained using a non-uniform discretization while the points connected with a dotted line represent results calculated with the equidistant discretization.
In both discretization schemes 1200 modes and cut-off value of $25k_0$ were used.
(b) The emission rate to radiation modes calculated with a fixed nanowire diameter $0.3\lambda$. The horizontal axis shows the number of discretization modes, and the legend shows the cut-off value of the wavenumber in units of $k_0$. The dotted red curves represent the results obtained using equidistant discretization while the solid lines are calculated using the non-uniform discretization. 
\label{fig:beta_gamma}\label{fig:beta_gamma_conv}}
\end{center}
\end{figure}

Similar to calculations represented in \cite{Claudon2013}, we consider a dipole emitter oriented along the wire axis in an infinitely long nanowire with $n_w = 3.45$ and surrounded by air. Figure \ref{fig:beta_gamma}(a) presents the $\beta$ factor and the emission rates to the fundamental guided mode ($HE_{11}$),
to the second guided mode ($HE{12}$),  and to the radiation modes, all normalized to the bulk emission rate (see Sec. \ref{sec:theory} \ref{sec:emission}) as functions of the nanowire diameter. While the rates calculated using both the equidistant and the non-uniform discretization schemes with 1200 modes and a cut-off value $25k_0$ agree well qualitatively, discrepancies are observed in the emission rate to radiation modes.  Figure \ref{fig:beta_gamma_conv}(b) shows a convergence investigation of the emission rate to radiation modes. 
We fix the nanowire geometry by setting the diameter as $0.3\lambda$, use both discretization schemes, and vary the cutoff value of the transverse wavenumber as well as the number of modes. The results show that only a slight improvement is achieved by increasing the cutoff from $20k_0$ to $25k_0$, while the results depend on the number of modes for small mode numbers and converge around 500. At high mode numbers and cut-off values the results converge to the same value.

\subsection{Reflection from semiconductor nanowire metal interface}

Finally, we investigate convergence of the method in a structure consisting of a nanowire standing on top of a metallic bulk layer. We calculate the reflection coefficient of the fundamental mode from a nanowire-metal interface similar to the setup investigated in \cite{Friedler08}. The refractive indices of the nanowire and metal are $n_w=3.5$ and $n_{\mathrm{Ag}}= \sqrt{-41 + 2.5\iim}$ at the wavelength $\lambda = 950$ nm. 

\begin{figure}[h]
\begin{center}
\includegraphics[width=0.95\columnwidth]
%{figures/reflection_nw_metal_all_edited.eps}
{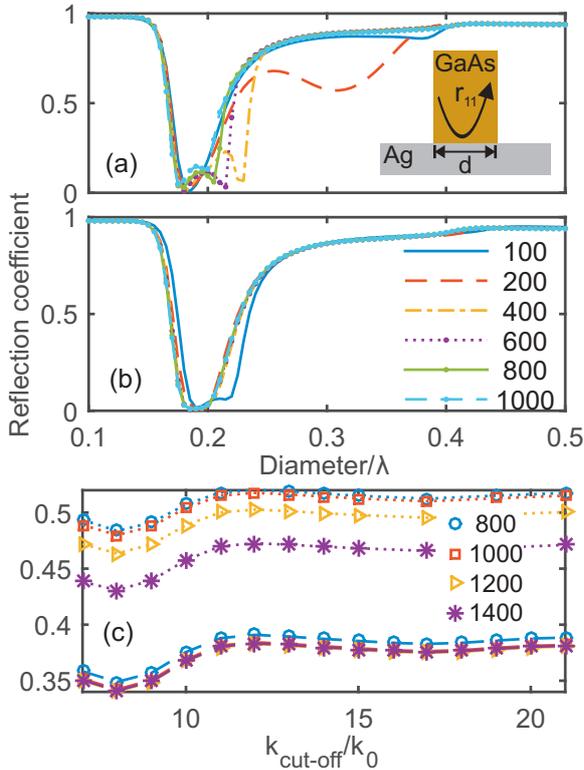}
\caption{
The reflection coefficient of the fundamental mode calculated using (a) an equidistant grid 
and (b) a nonuniform grid with varying number of modes (shown in the legend) and $k_{\mathrm{cut-off}}=20k_0$ as a function of the nanowire diameter. The wire and the metal have refractive indices of $n_w=3.5$ and $n_{\mathrm{Ag}}= \sqrt{41 + 2.5\iim}$, respectively, at wavelength $\lambda=950$ nm. 
(c)
The reflection coefficient of the fundamental mode using equidistant (dotted lines) and nonuniform (dashed lines) discretization and varying the cutoff of $k_m$ for a nanowire having diameter of $0.22\lambda$. The values of $k_m$ are chosen such that the $k_m$ is equidistantly/non-uniformly sampled up to value $2 n_w k_0$ ($n_w=3.5$) with $M$ shown in the legend. Then extra $k_m$ values are added according to the scheme when the cut-off value is increased.
\label{fig:R_grid_comp} \label{fig:R_kmax}}
\end{center}
\end{figure}

Figure \ref{fig:R_grid_comp} shows the calculated reflection coefficient as a function of the nanowire diameter using both (a) the equidistant sampling of the $k_{\perp}$ and (b) the nonuniform $k_{\perp}$ discretization with several different number of discretization modes. In the non-uniform discretization, the $k$-space values are sampled more densely close to $k_0$ as discussed in Sec. \ref{sec:disc}. With small wire diameter the reflection coefficients are essentially determined by the air-metal reflection  ($R_{\mathrm{Air-Ag}} \approx 0.98$) since in this limit the fundamental mode is mainly located in air. In contrast, in the limit of large nanowires the fundamental mode is primarily located into the GaAs wire ($R_{\mathrm{GaAs-Ag}} \approx 0.95$). Nevertheless, the figures show that faster convergence is obtained using the non-uniform discretization scheme instead of the equidistant $k$ discretization.

The reflection coefficients in Figs. \ref{fig:R_grid_comp}(a) and (b) are obtained for a fixed cut-off value. Next, we fix the geometry and study the effect of the cut-off value of $k_m$. We select a wire having diameter of $0.22\lambda$ since the reflection coefficients shown in the Fig. \ref{fig:R_grid_comp}(a)--(b) calculated with different discretization schemes and with varying number of modes  have large variations around this diameter. Reflection coefficients as functions of the cut-off value calculated using both discretization schemes with several different numbers of included modes are shown in Fig. \ref{fig:R_kmax}(c). The $k_m$ values are chosen such that when the cut-off is increased extra points are added to the original $k_m$ grid.  
The results show that the calculations converge around $5n_{\mathrm{w}}k_0$.

The convergence checks in the selected waveguide examples represented in Figs. \ref{fig:beta_gamma_conv} and \ref{fig:R_kmax} show convergence for the investigated waveguide sizes and structures. Although these examples do not guarantee the convergence of our method in all waveguide sizes and geometries, we expect our method to converge in various types and sizes of waveguides provided that geometry specific modifications to the discretization scheme are implemented.

\section{Conclusions}

We have demonstrated an open geometry Fourier modal method formalism relying on open boundary conditions and non-uniform $k$-space sampling. Due to the inherent open boundary conditions we avoid the artificial absorbing boundary conditions, that in some cases lead to numerical artifacts. 
We have tested the approach by investigating the dipole emission in a bulk, close to an interface, and in waveguide structures and by calculating the reflection coefficient of the fundamental waveguide mode for a nanowire-metal interface. Our simulations show that the calculations based on the open geometry Fourier modal method formalism indeed converge towards an open geometry limit when varying the cut-off and the number of modes and that the use of the non-uniform discretization scheme leads to a faster convergence of the simulations compared to using the conventional equidistant discretization. 
We expect that our new method will prove useful in accurate modeling of a variety of nanophotonic structures, for which the open boundaries are inherently difficult to describe. Also, extension of the formalism to three-dimensional Fourier modal method is straightforward, and could be used for accurate modeling of, for example, light emission in photonic crystal membrane waveguides \cite{Lecamp2007,MangaRao2007}.

%\section*{Funding}
\section*{Acknowledgments}
Support from the Danish Research Council for Technology and Production via the Sapere Aude project LOQIT (DFF - 4005-00370) is gratefully acknowledged.

%-----------------------------------------

\appendix

\section{Fourier-Bessel expansion in cylindrical coordinates} \label{app:rotsym}

The derivation of open BC method in rotationally symmetric case is outlined following the approach presented in \cite{Bonod2005}. We use cylindrical coordinates $(r,\phi,z)$. Since the considered structures are rotationally symmetric, the angular dependence is expanded using Fourier series  $\mathbf{E}(r,\phi, z) = \sum_{n=-\infty}^{\infty} \mathbf{E}_n(r,z) \exp(\iim n \phi )$. The contributions $\mathbf{E}_n(r,z)$ for different orders $n$ are decoupled, and it is thus possible to solve for each order individually. This advantage is exploited to reduce the 2D lateral eigenvalue problem to an effective 1D problem.

Using the Fourier expansion the time-harmonic Maxwell's equations
$\nabla \times \mathbf{E}(\mathbf{r}) = \iim\omega\mu_0 \mathbf{H}(\mathbf{r}) $
and
$\nabla \times \mathbf{H}(\mathbf{r}) = -\iim\omega \varepsilon(\mathbf{r}) \mathbf{E}(\mathbf{r}) $
 can be written component wise as
\begin{eqnarray}
  \frac{\partial}{\partial z}E_{\phi, n} &=& \frac{\iim n}{r}E_{z,n} -\iim \omega \mu_0 H_{r,n} \label{eq:me_Et} \\
\frac{\partial}{\partial z}E_{r,n}  &=& \frac{\partial}{\partial r}E_{z,n} +  \iim \omega \mu_0 H_{\phi, n} \label{eq:me_er}\\
\iim \omega \mu_0 H_{z,n} &=& \frac{\partial E_{\phi,n}}{\partial r}  + \frac{E_{\phi,n}}{r} - \frac{\iim n}{r} E_{r,n} \label{eq:me_Hz}\\
\frac{\partial H_{\phi,n}}{\partial z} &=& \frac{\iim n}{r}H_{z,n} + \iim \omega \varepsilon(r) E_{r,n} \label{eq:me_Ht}\\
\frac{\partial H_{r,n}}{\partial z}  &=& \frac{\partial H_{z,n}}{\partial r} -\iim \omega \varepsilon(r) E_{\phi,n} \label{eq:me_Hr}\\
-\iim \omega \varepsilon(r) E_{z,n} &=& \frac{\partial H_{\phi,n}}{\partial r} + \frac{H_{\phi,n}}{r} - \frac{\iim n}{r}H_{r,n} \label{eq:me_Ez} 
\end{eqnarray}

The Helmholtz equation for each Fourier component is given as  $\Delta  \mathbf{E}_n(r,z)\exp(\iim n\phi)+\omega^2\mu_0 \varepsilon(r) \mathbf{E}_n(r,z)\exp(\iim n\phi) = 0$ which in component wise reads as
\begin{eqnarray}
\Delta E_{r,n} - \frac{E_{r,n}}{r^2} - \frac{2\iim n}{r^2} E_{\phi,n}  + \omega^2\mu_0\varepsilon(r) E_{r,n} &=& 0 
\label{eq:BHrc}\\
\Delta E_{\phi,n} - \frac{E_{\phi,n}}{r^2} + \frac{2\iim n}{r^2} E_{r,n} + \omega^2\mu_0\varepsilon(r) E_{\phi, n} &=& 0 
\label{eq:BHtc}\\
\Delta E_z + \omega^2\mu_0\varepsilon(r) E_{z,n} &=& 0.
\label{eq:BHzc}
\end{eqnarray}
Equations (\ref{eq:BHrc}) and (\ref{eq:BHtc}) are of the form of Bessel differential equations and couple the radial and angular components of the Electric field. In order to simplify calculations these equations are de-coupled using the following notation
\begin{eqnarray}
%E_{n}^{+} = E_{\phi, n} + \iim E_{r, n},~ E_{n}^{-} = E_{\phi, n} - \iim E_{r, n} 
E_{n}^{\pm} = E_{\phi, n} \pm \iim E_{r, n}
\label{eq:E_pm}
\end{eqnarray} 
The transverse components of the Helmholtz equation can then be written as
\begin{eqnarray}
\Delta E_n^{+} - \frac{E_n^{+}}{r^2} + \frac{2\iim n}{r^2} E_n^{+}  + \omega^2\mu_0\varepsilon(r) E_n^{+} &=& 0 \\
\Delta E_n^{-} - \frac{E_n^{-}}{r^2} - \frac{2\iim n}{r^2} E_n^{-}  + \omega^2\mu_0\varepsilon(r) E_n^{-} &=& 0.  
%\label{eq:}
\end{eqnarray} 
These Bessel-differential equations have general solutions 
\begin{eqnarray}
\nonumber
E_n^{+} &=& E_{\phi, n} + \iim E_{r,n} = \int_{k_r=0}^{\infty} 2 c_n^E(k_r,z) J_{n-1}(k_r r) k_r \ud k_r \label{eq:Enp} \\ 
\\
\nonumber
E_n^{-} &=& E_{\phi, n} - \iim E_{r,n} = \int_{k_r=0}^{\infty} 2 b_n^E(k_r,z) J_{n+1}(k_r r) k_r \ud k_r \label{eq:Enm},\\
\end{eqnarray}
where $J_n$ is the Bessel function of the first kind having order of $n$.
For numerical calculations the above Bessel integrals are truncated as
$\int_{k_r=0}^{\infty}  k_r \ud k_r \longrightarrow \sum_{m=1}^{\mathrm{M}} k_m \Delta k_m$
and the Fourier series are truncated to $-N \le n \le N$.
The expansion are 
\begin{eqnarray}
\nonumber
E_r(r,\phi,z) &=& \iim \sum_{n=-N}^{N} \sum_{m=1}^{\mathrm{M}} k_m \Delta k_m \Big[ b_{n,m}^E(z) J_{n+1}(k_m r) 
\\&& \quad - c_{n,m}^E(z) J_{n-1}(k_m r) \Big]\exp(\iim n \phi) \label{eq:Er_exp} \\
\nonumber
E_{\phi}(r,\phi,z) &=& \sum_{n=-N}^{N} \sum_{m=1}^{\mathrm{M}} k_m \Delta k_m \Big[ b_{n,m}^E(z) J_{n+1}(k_m r) 
\\&& \quad + c_{n,m}^E(z) J_{n-1}(k_m r) \Big]\exp(\iim n \phi). \label{eq:Et_exp}
\end{eqnarray}
Equivalent equations are obtained for magnetic fields by substituting $c_n^E \rightarrow c_n^H$ and $b_n^E \rightarrow b_n^H$.
The $z$-components are obtained using the time-harmonic Maxwell's equations (\ref{eq:me_Hz}) and (\ref{eq:me_Ez}), above expansions,
and the derivation rules for Bessel functions as
\begin{eqnarray}
\nonumber
\iim \omega \mu_0 H_{z,n} 
&=& \sum_{m=1}^{\mathrm{M}} k_m^2 \Delta k_m [b_{n,m}^E - c_{n,m}^E]J_{n}(k_m r)\\
\\
\nonumber
-\iim \omega \varepsilon(r) E_{z,n} 
&=& \sum_{m=1}^{\mathrm{M}} k_m^2 \Delta k_m [b_{n,m}^H - c_{n,m}^H]J_{n}(k_m r).\\
\label{eq:Ez_bess}
\end{eqnarray}
To obtain expression for $E_{z,n}(r)$ we expand Eq. (\ref{eq:Ez_bess}) using $E_{z,n}=\sum_{m=1}^{\mathrm{M}} k_m \Delta k_m E_{z,n,m} J_{n}(k_m r)$, integrate both sides of Eq. (\ref{eq:Ez_bess}) with $\int_{r=0}^{\infty} \cdot r J_n(k_{m'}r) \ud r$ and use the orthogonality of Bessel functions. We then obtain expression for $E_{z,n,m}$  which is substituted to the expansion of $E_{z,n}$ giving
\begin{eqnarray} 
\nonumber
E_{z,n} &=& \frac{\iim}{ \omega } \sum_{m,m'=1}^{\mathrm{M}} 
  \left([\varepsilon]^{n,n}\right)_{m,m'}^{-1} k_{m'}  [b_{n,m'}^H - c_{n,m'}^H]  J_{n}(k_m r),
\\
\label{eq:Ez_exp}
\end{eqnarray}
where we have used short hand notation 
$[\varepsilon]_{m,m'}^{n,n} = \int_{r=0}^{\infty}\varepsilon(r)  J_{n}(k_m r) J_{n}(k_{m'} r) r \ud r$.

The expansion coefficients $b$ and $c$ are obtained representing the system as an eigenvalue problem by applying the differential method as follows. The $z$-dependence of the Maxwell's equations' expansion coefficients are written
as an eigenvalue problem
\begin{equation}
\frac{\ud \mathbf{f}_n(z)}{\ud z} = \mathbf{M}_n\mathbf{f}_n(z), \qquad n\in [-N,N],  
%\label{eq:}
\end{equation}
where $\mathbf{f} \in \mathbb{C}^{4M\times 1}$ and $\mathbf{M} \in \mathbb{C}^{4M\times 4M}$ are 
\begin{equation}
\mathbf{f}_n(z) = \left[ \begin{array}{l}
b_{n,m}^{E}(z) \\ c_{n,m}^{E}(z) \\ b_{n,m}^{H}(z) \\ c_{n,m}^{H}(z) 	
\end{array}\right]
\qquad %\mathrm{and} \qquad
\mathbf{M}_n = \left[ \begin{array}{ll}
M_{n,11} & M_{n,12} \\ M_{n,21} & M_{n,22} 	
\end{array}\right].
\end{equation}
Here the $z$-dependence is of the form $\exp(\iim \beta z)$.
% In the following it will be shown that matrices $M_{n,11}=M_{n,22} = 0$ so that
The derivatives of electric field expansion coefficients couple only to the magnetic field components and vice versa so that the propagation constants $\beta$ and expansion coefficients can be solved from eigenvalue problem 
\begin{eqnarray}
-\beta_n^2 \left[ \begin{array}{l} b_{n,m}^{E} \\ c_{n,m}^{E} \end{array}\right]
&=& M_{n,12}M_{n,21} \left[ \begin{array}{l} b_{n,m}^{E} \\ c_{n,m}^{E} \end{array}\right].
%
%\label{eq:}
\end{eqnarray}
The magnetic field expansion coefficients are obtained from the electric field ones by using matrix $M_{n,21}$. Equivalently, the eigenvalue problem can be written for magnetic field coefficients. 

Derivating definitions in Eq. (\ref{eq:E_pm}) with respect to z, substituting Maxwell's Eqs. (\ref{eq:me_Et})--(\ref{eq:me_Ez}), and using the orthogonality of Bessel functions
allows us to write for the electric field coefficients
\begin{widetext}
\begin{eqnarray}
%\nonumber
\frac{\ud \widetilde{b}_{n,m}^E}{\ud z}  
=  \hphantom{-} \omega\mu_0 \widetilde{b}_{n,m}^H - \frac{k_m}{2\omega}  \sum_{m'} \left([\varepsilon]_{m,m'}^{n,\widetilde{n}}\right)^{-1} k_{m'} [\widetilde{b}_{n,m'}^H - \widetilde{c}_{n,m'}^H] \label{eq:bE}&&\\
\frac{\ud \widetilde{c}_{n,m}^E}{\ud z}  
= -\omega\mu_0 \widetilde{c}_{n,m}^H - \frac{k_m}{2\omega}  \sum_{m'} \left([\varepsilon]_{m,m'}^{n,\widetilde{n}}\right)^{-1} k_{m'} [\widetilde{b}_{n,m'}^H - \widetilde{c}_{n,m'}^H] && 
%\label{eq:}
\end{eqnarray}
\end{widetext}
and for magnetic the field coefficients
\begin{eqnarray}
\nonumber
\frac{\ud \widetilde{b}_{n,m}^H}{\ud z} &=&  \frac{k_m^2}{2\omega \mu_0} (\widetilde{b}_{n,m}^E - \widetilde{c}_{n,m}^E ) \\
\nonumber
&&  +\iim \frac{1}{2} \omega k_m \int_{r=0}^{\infty} \varepsilon(r) E_{r,n}(r)  J_{n+1}(k_{m}r) r \ud r \\ 
\nonumber
&&    - \frac{1}{2}\omega k_m \int_{r=0}^{\infty} \varepsilon(r) E_{\phi, n}(r)  J_{n+1}(k_{m}r) r \ud r \\
\label{eq:bH}
\end{eqnarray}
\begin{eqnarray}
\nonumber
\frac{\ud \widetilde{c}_{n,m}^H}{\ud z} &=&  \frac{k_m^2}{2\omega \mu_0} (\widetilde{b}_{n,m}^E - \widetilde{c}_{n,m}^E ) \\
\nonumber
&&  +\iim \frac{1}{2} \omega k_m \int_{r=0}^{\infty} \varepsilon(r) E_{r,n}(r)  J_{n-1}(k_{m}r) r \ud r \\
\nonumber
&&    + \frac{1}{2}\omega k_m \int_{r=0}^{\infty} \varepsilon(r) E_{\phi, n}(r)  J_{n-1}(k_{m}r) r \ud r	\\
\label{eq:cH}
\end{eqnarray}
where $\widetilde{b}_{n,m}^{E} = k_m b_{n,m}^{E}$ and so on.
The integrals in Eqs. (\ref{eq:bH}) and (\ref{eq:cH}) involving $E_{\phi, n}(r)$ can be calculated using the direct rule\cite{Li96,Popov04} as 
\begin{eqnarray}
\nonumber
&&\int_{0}^{\infty} \varepsilon(r) E_{\phi, n}(r) J_{n\pm 1}(k_m r) r \ud r \\
\nonumber
&=& \sum_{m'=1}^{\mathrm{M}} k_{m'} \Delta k_{m'} \Big( [\varepsilon]_{m,m'}^{n\pm 1, n+1}  b_{n,m'}^E +  [\varepsilon]_{m,m'}^{n\pm 1, n-1}  c_{n,m'}^E \Big) \\
\end{eqnarray} 
while the integrals involving $E_{r, n}(r)$ are calculated using the inverse rule 
due to the discontinuities of $\varepsilon(r)$ and $E_{r, n}(r)$ as follows:
The electric field can be expanded using the electric displacement as
$E_{r, n}(r) = \sum_{m} k_m \Delta k_m\frac{1}{\varepsilon(r)} D_{r, n, m}^{\pm} J_{n\pm 1}(k_m r)$ which
has expansions components  given by $E_{r, n, m}^{\pm} = 
\int_{0}^{\infty} E_{r, n}(r) J_{n\pm 1}(k_m r) r \ud r$ so that
\begin{eqnarray}
\nonumber
E_{r, n, m'}^{\pm}  &=& \int_{0}^{\infty} E_{r, n}(r) J_{n\pm 1}(k_m r) r \ud r \\
&=&  \sum_{m} k_m \Delta k_m D_{r, n, m}^{\pm} \left[ \frac{1}{\varepsilon} \right]_{m',m}^{n\pm 1, n\pm 1} 
\end{eqnarray}
The expansion for electric displacement is obtained by inverting as
\begin{eqnarray}
\nonumber
 D_{r, n, m}^{\pm} &=& \sum_{m'}  \frac{1}{k_m \Delta k_m} \left( \left[ \frac{1}{\varepsilon} \right]^{n\pm 1, n\pm 1} \right)^{-1}_{m,m'} E_{r, n, m'}^{\pm}. \\
\label{eq:D_exp}
\end{eqnarray}
Solving $E_{r, n, m'}^{\pm}$ using Eq. (\ref{eq:Er_exp}) and substituting into Eq. (\ref{eq:D_exp}) leads to 
\begin{widetext}
\begin{eqnarray}
\nonumber
\iim k_m D_{r, n, m}^{+}  &=&  -\sum_{m'=1}^{\mathrm{M}} \frac{1}{k_{m'}\Delta k_{m}}
\left( \left[ \frac{1}{\varepsilon} \right]^{n+1,n+1} \right)^{-1}_{m,m'} k_{m'} b_{n,m'}^E \\
&&+ \sum_{m',m''=1}^{\mathrm{M}}  \frac{1}{k_{m'}\Delta k_{m}} \left( \left[ \frac{1}{\varepsilon} \right]^{n+1,n+1} \right)^{-1}_{m,m'} 
%\\ \nonumber && \qquad 
\times 
k_{m'} \Delta k_{m''} \left[ \Psi\right]_{m',m''}^{n+1, n-1}  k_{m''}c_{n,m''}^E \\
\nonumber
\iim k_m D_{r, n, m}^{-} &=&   
-\sum_{m',m''=1}^{\mathrm{M}}  \frac{1}{k_{m'}\Delta k_{m}} \left( \left[ \frac{1}{\varepsilon} \right]^{n-1,n-1}\right)^{-1}_{m,m'}
%  \\ \nonumber  && \qquad 
\times k_{m'}\Delta k_{m''} \left[ \Psi\right]_{m',m''}^{n-1, n+1} k_{m''} b_{n,m''}^E \\
&&+ \sum_{m'=1}^{\mathrm{M}} \frac{1}{k_{m'}\Delta k_{m}} \left( \left[ \frac{1}{\varepsilon} \right]^{n-1,n-1}\right)^{-1}_{m,m'}  k_{m'}c_{n,m''}^E, 
\end{eqnarray}
\end{widetext}
where the following notations were used 
\begin{eqnarray}
\nonumber
\left[\frac{1}{\varepsilon}\right]_{m,m'}^{n,n} &=& \int_{r=0}^{\infty} \frac{1}{\varepsilon(r)}  J_{n}(k_m r) J_{n}(k_{m'} r) r \ud r \\
\\
\nonumber
~[\Psi]_{m,m'}^{n\pm 1,n\mp 1} &=& \int_{r=0}^{\infty}  J_{n\pm 1}(k_m r) J_{n\mp 1}(k_{m'} r) r \ud r.
\\
\end{eqnarray}

%******************************************
% Bibliography
%\bibliography{References_OpenFMM}
\input{OpenFMM_arxiv.bbl}

\end{document}

%% file: OpenFMM_arxiv.bbl
%merlin.mbs apsrev4-1.bst 2010-07-25 4.21a (PWD, AO, DPC) hacked
%Control: key (0)
%Control: author (8) initials jnrlst
%Control: editor formatted (1) identically to author
%Control: production of article title (-1) disabled
%Control: page (0) single
%Control: year (1) truncated
%Control: production of eprint (0) enabled
%